\documentclass[prd,aps,showpacs,amsfonts,eucal]{revtex4}
\usepackage{graphicx}

\begin{document}

\bibliographystyle{prsty} 

\title{
Late time tails from momentarily stationary, compact initial 
data in Schwarzschild spacetimes}
\author{Richard H.~Price}
\affiliation{Department of 
Physics and Astronomy and Center for Gravitational Wave Astronomy, The
University of Texas at Brownsville, Brownsville, Texas, 78520}
\author{Lior M.~Burko}\affiliation{Department of Physics
and Astronomy, Bates College, Lewiston, Maine 04240}
\date{\today}

\begin{abstract}
\begin{center}
{\bf Abstract}
\end{center}
An $\ell$-pole perturbation in Schwarzschild spacetime generally falls
off at late times $t$ as $t^{-2\ell-3}$.  It has recently been pointed
out by Karkowski, \'{S}wierczy\'{n}ski and Malec, that for initial data
that is of compact support, and is initially momentarily static, the
late-time behavior is different, going as $t^{-2\ell-4}$. By considering
the Laplace transforms of the fields, we show here why the momentarily
stationary case is exceptional. We also explain, using a time-domain
description, the special features of the time development in this
exceptional case.
\end{abstract}

\maketitle
\section{Introduction}\label{sec:intro} 
The perturbations in Schwarzschild spacetime radiate into the horizon
and out to future null infinity (Scri+), so that at any fixed position in
Schwarzschild coordinates, the perturbation falls off in time $t$. It has
long been known that the fall-off is an inverse power-law in time of the
form $t^{-n}$.  For a multipole perturbation of multipole index $\ell$,
the value of the power law index $n$ can take several forms. In the case
of a perturbation that, at large radius, has the asymptotic form of a
static multipole, $n=2\ell+2$. This case is of particular astrophysical
interest since it describes the fate of initial multipoles coupled to a
star that has been stationary, but undergoes gravitational collapse to a
black hole \cite{price}. This result has been confirmed, for example, by
Cunningham, Price, and Moncrief \cite{cpm}, and more recently by
Baumgarte and Shapiro, in the context of the collapse of magnetized
neutron stars \cite{baumshap}.

In the case of an initial moment that is asymptotically static, the
initial field at large radius is the limiting factor in the rate at
which the field falls off. The fall-off is faster if the initial data
has compact support.  The rule in this case is $n=2\ell+3$
\cite{price,manyrefs,barack}. This result was found numerically
also for fully-nonlinear spherical collapse of a scalar field
\cite{gpp,bo}. Recently, however, Karkowski, \'{S}wierczy\'{n}ski and
Malec \cite {karkswiermalec}, hereafter KSM, presented numerical evidence
that if the data are momentarily stationary as well as being of compact
support, then $n=2\ell+4$. Here we will explain, from two different
points of view, why the momentarily static case is an exception. 

Perturbations of spherically symmetric black holes can be decomposed
into multipoles, and each multipole moment $\Psi(x,t)$ satisfies an
equation of the form
\begin{equation}\label{geneq} 
\frac{\partial^2\Psi}{\partial x^2}-\frac{\partial^2\Psi}{\partial t^2}
-V(x)\Psi=0\,.
\end{equation}
In the specific case of perturbations of a Schwarzschild spacetime of
mass $M$, the variable $t$ is the usual Schwarzschild time coordinate
and $x$ is the `tortoise' coordinate, related to the Schwarzschild
radial coordinate $r$ by $x\equiv r_*=r+2M\ln{(r/2M-1)}$. Here we use
units in which $G=1=c$; we choose $M=1$ without loss of generality, so
that $t$ and
$x$ are dimensionless.  If $\Psi$ represents odd-parity gravitational
perturbations, then the potential $V(x)$ is the Regge-Wheeler \cite{RW}
potential; if $\Psi$ represents even-parity gravitational perturbations,
$V(x)$ is the Zerilli \cite{zerilli} potential.  For even- or odd-parity
electromagnetic perturbations, or for scalar perturbations, $V(x)$ has
a somewhat different form.  It will be convenient here for us not to
specify at the outset just what particular form $V(x)$ takes. We will
require only that $V(x)$ falls off sharply as $x\rightarrow-\infty$\,
and that $V(x)=\ell(\ell+1)x^{-2}\{1+{\cal O}[\ln(x)/x]\}$, for large $x$.

In the next section we work with the Laplace transform of $\Psi(x,t)$,
and relate the Laplace transform to an integral over the initial
data. In principle, the form of the late-time tails of the
perturbations can be extracted from the analytic details of Green
functions in Laplace or Fourier space, as others have
shown \cite{leaver,laplace}. Indeed, as shown by Leaver \cite{leaver}, the
late
time tails will be a sum of a $t^{-2\ell -3}$ tail and a $t^{-2\ell
-4}$ tail, the latter term arising from time-symmetric initial data.
But we can avoid such complications. If one accepts that $n=2\ell+3$ is
the result for generic initial data of compact support, it turns out that
an immediate consequence is that $n$ must be $2\ell+4$ if the initial data
are momentarily stationary.

The late-time tails are usually thought of as a result of backscatter    
of radiation by the potential $V(x)$ at large radius. Though the proof
in Sec.~\ref{sec:laplace} is definitive, it does not explain how, in
the scattering picture, the momentarily stationary initial data     
are exceptional. In Sec.~\ref{sec:backscatter} we provide a heuristic
explanation by showing that for the momentarily stationary case, the
initial data result in two outgoing pulses that, in a sense,
cancel each other. 

The exceptional behavior of time-symmetric initial data can suggest
the following paradox: Take the effective potential to be
that of a Schwarzschild spacetime, but truncate it below a certain
value of the Regge-Wheeler `tortoise' coordinate $r_*$, and take this
truncation to be at a large negative value of $r_*$. 
For the generation of tails, such a truncated potential is expected
to be an excellent approximation to the full Schwarzschild potential,
because the Schwarzschild potential drops off exponentially with $r_*$
for large and negative values of $r_*$, and as is well known, it is
only the form of the effective potential at large distances (large and
positive values of $r_*$) that is important for the tails problem in
Schwarzschild.
Consider first initial data of an outgoing pulse of compact support to
the ``left'' (more negative $r_*$ side) of the truncated potential, so
that the initial pulse is fully located in the region of zero
potential.  One could expect the tail in this case to be given by
$t^{-2\ell -3}$, since the initial outgoing pulse is generic
time-asymmetric initial data. Consider next the same situation, but
this time with an initially momentarily static pulse of twice the
amplitude of the initially outgoing pulse we previously
considered. The compact initial pulse is in a region of spacetime with
zero potential, and therefore will immediately split into outgoing and
incoming pulses. The latter is never heard from again; it travels to
the left in a region of zero potential, and therefore never
scatters. The pulse traveling to the right is identical to the
situation we considered above. However, in this case, based on the
prediction of KSM, the tails should be given by $t^{-2\ell -4}$. How
can we explain this paradox, and what is the correct form of
the tail in this situation? We conclude this paper by resolving this
conflict of predictions.

\section{Relation of tails to initial data}\label{sec:laplace} 
We now follow the approach used by several authors \cite{leaver,laplace}, 
and introduce the Laplace transform $\psi(x,s)$ of $\Psi(x,t)$
through
\begin{equation}
{\cal L}[\Psi(x,t)]\equiv\psi(x,s)=\int_0^\infty e^{-st}\Psi(x,t)\,dt
\end{equation}
and the inverse
\begin{equation}
{\cal L}^{-1}[\psi(x,s)]\equiv\Psi(x,t)=\frac{1}{2\pi i}\int_\Gamma
e^{st}\psi(x,s)\,ds\,,
\end{equation}
where $\Gamma$ is a vertical contour in the right half of the complex
$s$ plane. With the relation ${\cal L}(\partial\Psi/\partial
t)=-\Psi(x,t=0)+s\psi(x,s)$, and its extension to second time derivatives, 
we write the Laplace transform of Eq.~(\ref{geneq}) as
\begin{equation}\label{genlap} 
\frac{\partial^2\psi}{\partial x^2}-\left[s^2+V(x)\right]\psi=S(x,s)
=-s\Psi_0(x)-\dot{\Psi}_0(x)\,.
\end{equation}
Here $\Psi_0$ and $\dot{\Psi}_0$ are, respectively, the initial
($t=0$) value of $\Psi(x,t)$ and the initial value of
$\partial\Psi(x,t)/\partial t$.

To solve Eq.~(\ref{genlap}), we again follow the approach of several
authors \cite{leaver,laplace}; we define homogeneous solutions $y_L(x,s)$
and $y_R(x,s)$
of Eq.~(\ref{genlap}) that, respectively, represent waves moving inward through
the horizon, and outward at spatial infinity:
\begin{equation}
y_L\stackrel{x\rightarrow -\infty}{\sim}e^{sx}
\quad\quad\quad
y_R\stackrel{x\rightarrow+\infty}{\sim}e^{-sx}\,.
\end{equation}
The Green function can be constructed in the usual way from 
$y_L$ and $y_R$, and the solution to Eq.~(\ref{genlap}) is given by 
\begin{equation}\label{gensoln} 
\psi(x,s)=\frac{1}{W(s)}\left[
y_R(x,s)\int_{-\infty}^xy_L(x',s) S(x',s)\,dx'
+
y_L(x,s)\int_x^
{\infty}y_R(x',s) S(x',s)\,dx'
\right]\, ,
\end{equation}
$W(s)$ being the Wronskian determinant of the homogeneous solutions. 
We next use the form of $S(x,s)$ in Eq.~(\ref{genlap})
to write Eq.~(\ref{gensoln}) as
\begin{equation}
\psi(x,s)=\psi_0(x,s)+\dot{\psi}_0(x,s)\,,
\end{equation}
where 
\begin{eqnarray}
\psi_0(x,s)
&=&\frac{-s}{W(s)}\left[
y_R(x,s)\int_{-\infty}^xy_L(x',s) \Psi_0(x')
\,dx'
+
y_L(x,s)\int_x^
{\infty}y_R(x',s) \Psi_0(x')\,dx'
\right]\label{psi0ofs}  \\
\dot{\psi}_0(x,s)
&=&\frac{-1}{W(s)}\left[
y_R(x,s)\int_{-\infty}^xy_L(x',s) \dot{\Psi}_0(x')
\,dx'
+
y_L(x,s)\int_x^
{\infty}y_R(x',s) \dot{\Psi}_0(x')\,dx'
\right]\,.\label{psi0dotofs}
\end{eqnarray}

Now let us suppose that $\dot{\Psi}_0(x)$ and ${\Psi}_0(x)$ are
arbitrary (bounded) functions of compact support, and  for every
choice of these functions, except perhaps the choice
$\dot{\Psi}_0(x)=0$, let us suppose that the fields at any value
of $x$ fall off as $t^{-2\ell-3}$. From this we conclude that 
for any bounded $f(x)$ of compact support
the expression
\begin{equation}
F(x,s)=\frac{-1}{W(s)}\left[
y_R(x,s)\int_{-\infty}^xy_L(x',s) f(x')
\,dx'
+
y_L(x,s)\int_x^
{\infty}y_R(x',s) f(x')\,dx'
\right]\,,
\end{equation}
gives the Laplace transform of a function that falls off in time no
slower than $t^{-2\ell-3}$. Now note that $sF(x,s)$ is the Laplace
transform of the time derivative of this function, and that the time
derivative will fall off as $t^{-2\ell-4}$. We can therefore conclude
that if the generic late time behavior is $t^{-2\ell-3}$, then
$\dot{\psi}_0(x,s) $ is the transform of a function that falls off as
$t^{-2\ell-3}$ and
${\psi}_0(x,s) $ is the transform of a function that falls off as
$t^{-2\ell-4}$. In the exceptional case that the initial data
is momentarily stationary, 
$\dot{\psi}_0(x,s)$ vanishes, and the late time behavior is 
$t^{-2\ell-4}$.

\section{The backscatter of momentarily stationary compact initial
data}\label{sec:backscatter}

To explain the $n=2\ell+4$ tails we shall use the general heuristic
framework developed in Refs.~\cite{ori,barack}: consider that there is 
a background problem, with a zero-order potential
\begin{equation}
V_0^\ell\equiv\left\{
\begin{array}{cl}
\ell(\ell+1)/x^2\quad&x>x_{0}\\
0&x<x_{0}
\end{array}
\right.\,.
\end{equation}
We will consider the remainder of the potential to be a perturbation,
so that $V(x)=V_0^\ell +\epsilon\, \delta V$.  The $\epsilon$ is
an accounting device so that we carry out a sort of perturbative
analysis.

The idea of this division of the potential into a background part and
a perturbation is that the background is a pure centrifugal potential
that cannot produce long-lived radiative tails. The tails, therefore,
must be due to $\delta V$. The (scalar field) monopole case is somewhat
awkward, since
the centrifugal potential vanishes.  What we really need though is
some ``edge'' at some $x_0$. Barack \cite{barack} discusses the
possibility of using a delta function for this purpose, but we need
not be specific. Despite its awkward feature, we shall rely heavily on
the monopole case. This is not solely because the slowly decaying
$\ell=0$ tails are the easiest to compute to long times. More
important, the description of backscatter for the $\ell=0$ case, lacks
technical complications of higher order multipoles. To focus on the
essential ideas of backscatter, we shall therefore confine ourselves
to $\ell=0$.  The extension to higher $\ell$ is straightforward.

We shall confine ourselves to descriptions to first order in
$\epsilon$. This, {\it sensu stricto}, is not correct but we believe
that the fundamental picture that comes out of that first-order
analysis {\it is} correct.  Strong evidence for this is the numerical
accuracy (illustrated below in Fig.~\ref{fig:area_time}) of a
prediction coming from this picture.

As in Ref.~\cite{barack}, we shall introduce 
advanced time $v$ and retarded time $u$ by
\begin{equation}
v=t+x\quad\quad u=t-x\,,
\end{equation}
and we shall focus attention, not on tails at ${\cal I}^+$ (that is,
$t\rightarrow\infty$ at constant $x$), but rather at Scri+ (that is,
$v\rightarrow\infty$ at constant $u$).  It can be
shown \cite{price,barack} that the tails at Scri+ and at ${\cal
I}^+$ are tightly connected. If the former is $u^{-m}$ then the latter
is $t^{-m-\ell-1}$. For the monopole case, then, we need to show that
for generic initial data of compact support, the tail at Scri+ falls
off as $u^{-2}$, while for momentarily stationary initial data of
compact support the tail has the form $u^{-3}$.

To simplify some statements in our analysis, we will not deal with the
monopole potential {\em per se}, but rather, shall take our potential
to be strictly zero for $x<x_0$, and to be $1/x^3$ for $x>x_0$.
We now write Eq.~(\ref{geneq}), to first order in $\epsilon$ as
\begin{equation}\label{Psiuv} 
\Psi_{(1),uv}=-\frac{1}{4}\,\epsilon\;\delta V \Psi_0\,.
\end{equation}
Let us suppose that the zeroth order solution is an 
outgoing pulse $\Psi=F(u)$ of compact support. 
Following the steps in Ref.~\cite{barack}, and making the same
approximations,  we get
\begin{equation}
\Psi_{(1)}=-\epsilon\,\frac{1}{4}\int_{-\infty}^u F(u')\,du'
\int_{u+2x_0}^\infty dv'\,\delta V(x')\,.
\end{equation}
When we use our special form of the potential $\delta V=1/x^3$, 
this becomes
\begin{equation}\label{tailamp} 
\Psi_{(1)}=-\epsilon\,\frac{1}{4}\int_{-\infty}^u F(u')
du'\,\int_{u+2x_0}^\infty dv'\, \frac{8}{(v'-u')^3 } \approx
-\frac{\epsilon}{u^2 }\int_{-\infty}^{\infty} F(u') du' \,.  
\end{equation}
We have
used the same approximations here as those of Ref.~\cite{barack}. In
particular, we have assumed that the value of $u$ to which this tail
result is to be valid satisfies $u\gg u'$, where $u'$ is any point 
in the support for $F$ \cite{accuracy}.

\begin{figure}[ht] 
\includegraphics[width=2.5in]{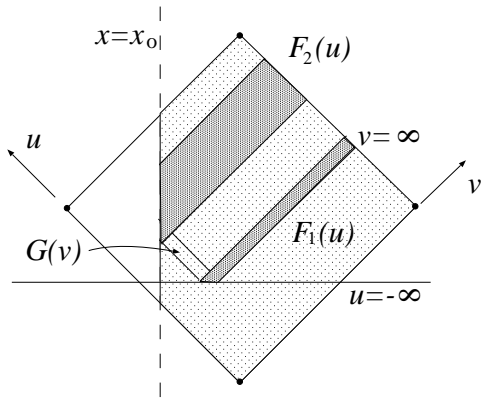} \caption{The evolution
of momentarily stationary initial data of compact support. The 
figure shows the zeroth order pulses $F_1(u)$, $G(v)$, and $F_2(u)$
described in the text.
\label{fig:initstat}}
\end{figure}

We now consider the case of initial data that are momentarily
stationary. The zero-order solutions for such initial data will
immediately ``split'' into an ingoing pulse and an outgoing pulse.
These are labeled as $F_1(u)$ and $G(v)$ in Fig.~\ref{fig:initstat}.
(Note that $v$ has been conformally rescaled in the figure to bring
Scri+ to a finite location.) In the special case of momentarily
stationary initial data the ingoing and outgoing zero-order pulses
will be related by $G(\xi)=F_1(-\xi)$.
The ``edge'' at $x=x_0$ is a zero-order
feature, so the ingoing $G(v)$ will undergo partial reflection at $x_0$
and will generate a second outgoing pulse $F_2(u)$.

For there to be no $u^{-2}$ tail at Scri+ (and hence no $t^{-3}$ tail
at ${\cal I}^+$) it must be the case that $\int_{-\infty}^
{+\infty}
\left[F_1(u)+F_2(u)
\right]\,du=0$. We have numerically checked a large number of 
examples, with different potentials, and different momentarily
stationary initial data. In all cases we have found that the ``cross
section'' (i.e.\,, the $u$ integral) of the reflected pulse $F_2(u)$
is opposite in sign to $F_1(u)$ and to numerical accuracy is equal
in magnitude. Since $G(\xi)=F_1(-\xi)$, this is equivalent to 
\begin{equation}\label{GplusF} 
\int G(v)\,dv+\int F_2(u)\,du=0\,.
\end{equation}

\begin{figure}
\input epsf
\centerline{ \epsfxsize 7.0cm
\epsfbox{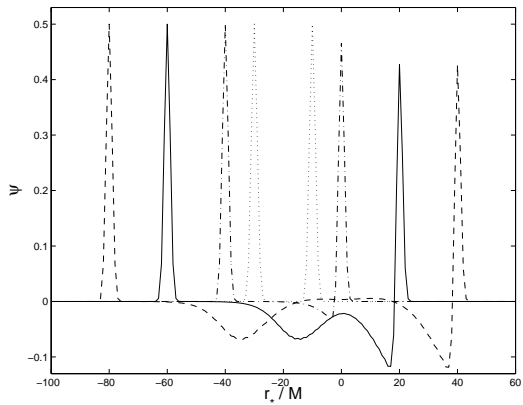}}
\caption{The field at four different value of the time. The initial data
is that of a time-symmetric field centers at $r_*=-20M$, and the field is
shown for $t=10M$ (dotted), $t=20M$ (dash-dotted), $t=40M$ (solid), and
$t=60M$ (dashed curve).} 
\label{fig:time_slices}
\end{figure}

In practice, we integrated at $t={\rm const}$ over the outgoing part of
the field. To have a numerically zero integral, the reflected field $F_2$
must have the opposite sign to the initial field $F_1$. In
Fig.~\ref{fig:time_slices} we show the field at different values of the
time
as a function of the `tortoise' coordinate. The figure shows the field
soon after the time-symmetric initial data split into outgoing
and incoming fields. As the outgoing field arrives at the peak of the
effective potential the field scatters, and part is reflected toward the
left with the opposite sign (and never heard from again), and a field of the
opposite sign continues to move toward the right, following the main
pulse. That is, the outgoing field is composed of the prompt field, and a
broadened field of the opposite sign. It is the integral of the combined
outgoing field that we calculate, and the result is shown in
Fig.~\ref{fig:area_time}. In practice, we compute the integral only for
positive values of $r_*$, to capture only the contributions from the
outgoing field. However, at and near $r_*=0$ there is no clear separation
between outgoing and incoming fields, and the field there is not strictly
zero for finite values of time. Because of the contributions from the
neighborhood of the peak of the effective potential, the integral does not
vanish at finite values of time. However,  the ``area'' between the field
and the horizonal axis drops with time, and as $t\to\infty$ the integral
approaches zero, like $t^{-2}$.

\begin{figure}
\input epsf
\centerline{ \epsfxsize 7.0cm
\epsfbox{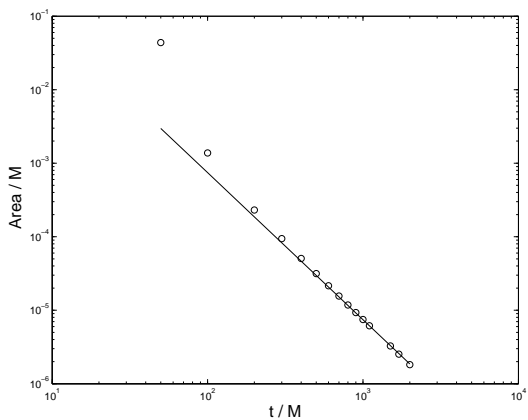}}
\caption{The ``area'' integral of the outgoing field vs.~the time. The
circles represent the numerically integrated area, and the solid reference
line is $7.44/t^2$. The errors in the numerical data are smaller than
$1\%$.} 
\label{fig:area_time}
\end{figure}

This should not be misinterpreted as total reflection of the ingoing
pulse $G(v)$, in the sense of total reflection of energy in the
waves. Such a statement about reflection refers to a quantity
quadratic in the wave pulse; the ``reflected'' pulse $F_2(u)$ is
generally quite different in shape from the ingoing pulse $G(v)$, so
Eq.~(\ref{GplusF}) is very different from a claim of total energy
reflection.
\begin{figure}[ht] 
\includegraphics[width=2.5in]{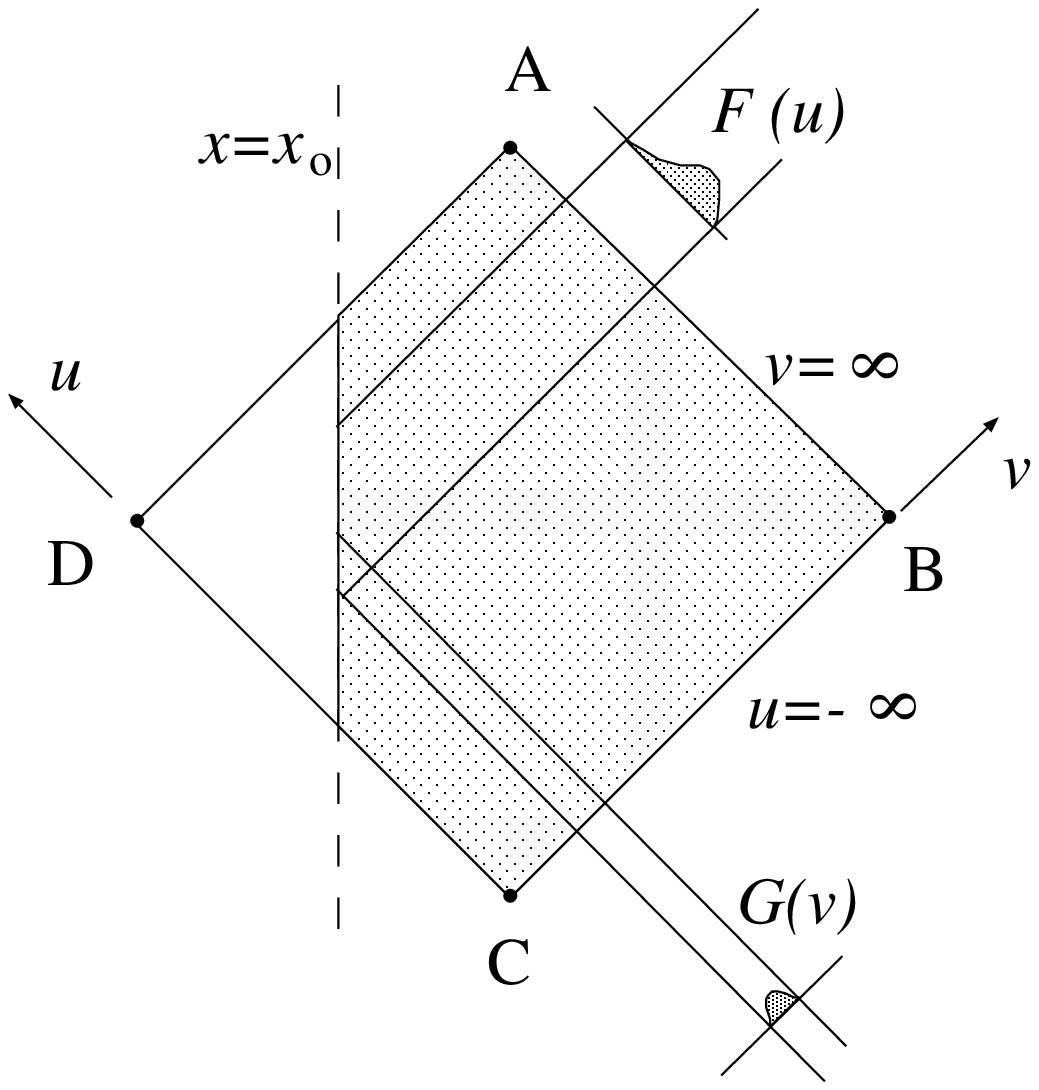} \caption{ 
Reflection of an ingoing zeroth-order wave at the 
edge of the potential.
\label{fig:intoout}} 
\end{figure}

The relationship in Eq.~(\ref{GplusF}) can be said to be the
explanation for initially stationary initial data being a special
case, and therefore of some importance. Though numerical verification
of this relationship is its ultimate justification, it is interesting
that there is a heuristic argument for Eq.~(\ref{GplusF}) that helps
us to understand it. Figure \ref{fig:intoout} shows a zeroth-order
ingoing wave reflecting off the edge at $x=x_0$. For this situation,
let us integrate the relationship in Eq.~(\ref{Psiuv}) over the range
of $u,v$ shown in Fig.~\ref{fig:intoout} as the rhombus with
vertices $A$,$B$,$C$,$D$.
The right hand side is clearly of order $\epsilon$. We make this explicit
by writing
\begin{equation}\label{intuvformeps} 
\int\int\Psi_{(1),uv}\,du\,dv=-\frac{1}{4}\epsilon\,\int\int\delta
V\Psi_{(0)}
\,du\,dv\,.
\end{equation}
The integral on the left can immediately be evaluated:
\begin{equation}\label{PsiABCD} 
\int\int\Psi_{(1),uv}\,du\,dv=\Psi_{(1)A}+\Psi_{(1)C}-\Psi_{(1)B}-\Psi_{(1)D}.
\end{equation}
From causality we have
\begin{equation}
\Psi_{(1)C}=\Psi_{(1)D}=0\,.
\end{equation}
By taking point $A$ at sufficiently large $u$, we can make
$=\Psi_{(1)A} $ arbitrarily small. Point $B$ is is at $I^+$. The
scattering of the ingoing pulse to  $I^+$ is zero, so
$\Psi_{(1)B} =0$.

We conclude that the left hand side of Eq.~(\ref{intuvformeps})
vanishes, and this means that the integral on the right of
Eq.~(\ref{intuvformeps}) must vanish. We can break the right hand side
integral  into the contributions due to the zeroth order ingoing
pulse $G(v)$ and the zeroth order outgoing pulse $F(u)$. For the ingoing
pulse
\begin{equation}
\int\int
\delta V G(v)\;du\,dv=2\int\int \delta V G(v)\;dx\,dv=
2\int_{r_0}^\infty
\delta V(x)\,dx\;\int G(v)\,dv\,.
\end{equation}
If we add the outgoing contribution, we find that
\begin{equation}\label{sumvanish} 
2\int_{x_0}^\infty
\delta V(x)\,dx\;\left[\int G(v)\,dv+\int F(u)\,du\;\right]
\end{equation}
must vanish, and hence we have given a heuristic 
explanation for Eq.~(\ref{GplusF}).

\begin{figure}
\input epsf
\centerline{ \epsfxsize 7.0cm
\epsfbox{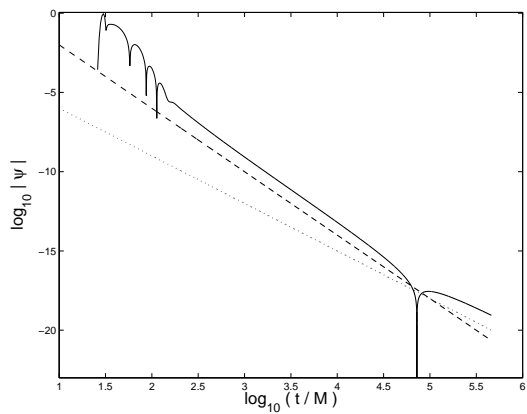}}
\caption{A spherical scalar field in Schwarzschild. The initial data are
for an outgoing compact pulse centered at $r_*/M=-40$. The solid line is the
field, the dashed reference line is proportional to $t^{-4}$, and the
dotted reference line is proportional to $t^{-3}$.}
\label{fig:pulse_left}
\end{figure}

We are now in a position to revisit the  paradox we described
in the Introduction. There is a fundamental, although subtle,
difference between the case of the truncated potential and the true
Schwarzschild potential: In the latter case there is a small overlap
of the initial data and the rapidly decreasing effective
potential. Because of this overlap, there is no {\em exact}
equivalence between outgoing initial data and time-symmetric initial
data of twice the amplitude. Because the overlap is small, we expect
the field from an initially outgoing pulse to fall off as $t^{-2\ell
-4}$ at intermediate times. However, the small overlap implies that
the true late time behavior will be $t^{-2\ell -3}$. This situation is
demonstrated in Fig.~\ref{fig:pulse_left}, which shows the results for
an initially outgoing field that is located to the left of the peak of
the effective potential, for a spherically symmetric scalar field. At
intermediate times the field clearly falls off like $t^{-4}$, but the
asymptotic fall off is $t^{-3}$, as expected.

\section*{Acknowledgment} 
We thank Lior Barack for useful comments. 
We gratefully acknowledge the support of the
National Science Foundation  under grant
PHY0244605, originally awarded to the University of Utah,
where this work was started.





\begin{thebibliography}{99}
\bibitem{accuracy} 
Numerical experiments have confirmed that the amplitude 
of the $u^{-2}$ tail at Scri+ is proportional to the integral
of the outgoing pulse, but the numerical values found are 
smaller than those in Eq.~(\ref{tailamp}) by a factor of 
around 0.88. This may be due to the contributions of higher
order scattering.
\end{thebibliography}
\end{document}